\documentclass{webofc}
\usepackage[varg]{txfonts}   
\usepackage{amssymb, amsmath}
\usepackage{hyperref}
\usepackage{textcomp}
\begin{document}
\title{IAS/CEA Evolution of Dust in Nearby Galaxies (ICED): the spatially-resolved dust properties of NGC4254}
    \author{\lastname{L.~Pantoni}\inst{\ref{CEA},\ref{IAS}}\fnsep\thanks{\email{lara.pantoni@cea.fr}}
              \and  R.~Adam \inst{\ref{OCA}}
              \and  P.~Ade \inst{\ref{Cardiff}}
              \and  H.~Ajeddig \inst{\ref{CEA}}
              \and  P.~Andr\'e \inst{\ref{CEA}}
              \and  E.~Artis \inst{\ref{LPSC},\ref{Garching}}
              \and  H.~Aussel \inst{\ref{CEA}}
              \and  M.~Baes  \inst{\ref{Belguim}}
              \and  A.~Beelen \inst{\ref{LAM}}
              \and  A.~Beno\^it \inst{\ref{Neel}}
              \and  S.~Berta \inst{\ref{IRAMF}}
              \and  L.~Bing \inst{\ref{LAM}}
              \and  O.~Bourrion \inst{\ref{LPSC}}
              \and  M.~Calvo \inst{\ref{Neel}}
              \and  A.~Catalano \inst{\ref{LPSC}}
              \and  M.~De~Petris \inst{\ref{Roma}}
              \and  F.-X.~D\'esert \inst{\ref{IPAG}}
              \and  S.~Doyle \inst{\ref{Cardiff}}
              \and  E.~F.~C.~Driessen \inst{\ref{IRAMF}}
              \and  G.~Ejlali \inst{\ref{Tehran}}
              \and  F.~Galliano \inst{\ref{CEA}}
              \and  A.~Gomez \inst{\ref{CAB}} 
              \and  J.~Goupy \inst{\ref{Neel}}
              \and  A.~P.~Jones \inst{\ref{IAS}}
              \and  C.~Hanser \inst{\ref{LPSC}}
              \and  A.~Hughes \inst{\ref{IRAP}}
              \and  S.~Katsioli \inst{\ref{Athens_obs},\ref{Athens_univ}}
              \and  F.~K\'eruzor\'e \inst{\ref{Argonne}}
              \and  C.~Kramer \inst{\ref{IRAMF}}
              \and  B.~Ladjelate \inst{\ref{IRAME}} 
              \and  G.~Lagache \inst{\ref{LAM}}
              \and  S.~Leclercq \inst{\ref{IRAMF}}
              \and  J.-F.~Lestrade \inst{\ref{LERMA}}
              \and  J.~F.~Mac\'ias-P\'erez \inst{\ref{LPSC}}
              \and  S.~C.~Madden \inst{\ref{CEA}}
              \and  A.~Maury \inst{\ref{CEA}}
              \and  P.~Mauskopf \inst{\ref{Cardiff},\ref{Arizona}}
              \and  F.~Mayet \inst{\ref{LPSC}}
              \and  A.~Monfardini \inst{\ref{Neel}}
              \and  A.~Moyer-Anin \inst{\ref{LPSC}}
              \and  M.~Mu\~noz-Echeverr\'ia \inst{\ref{LPSC}}
              \and  A.~Nersesian \inst{\ref{Belguim}}
              \and  D.~Paradis \inst{\ref{IRAP}}
              \and  L.~Perotto \inst{\ref{LPSC}}
              \and  G.~Pisano \inst{\ref{Roma}}
              \and  N.~Ponthieu \inst{\ref{IPAG}}
              \and  V.~Rev\'eret \inst{\ref{CEA}}
              \and  A.~J.~Rigby \inst{\ref{Leeds}}
              \and  A.~Ritacco \inst{\ref{ENS}, \ref{INAF}}
              \and  C.~Romero \inst{\ref{Pennsylvanie}}
              \and  H.~Roussel \inst{\ref{IAP}}
              \and  F.~Ruppin \inst{\ref{IP2I}}
              \and  K.~Schuster \inst{\ref{IRAMF}}
              \and  A.~Sievers \inst{\ref{IRAME}}
              \and  M.~W.~S.~L.~Smith \inst{\ref{Cardiff}}
              \and  F.~S.~Tabatabaei \inst{\ref{IPM}}
              \and  J.~Tedros \inst{\ref{IRAME}}
              \and  C.~Tucker \inst{\ref{Cardiff}}
              \and  E.~M.~Xilouris \inst{\ref{Athens_obs}}
              \and  R.~Zylka \inst{\ref{IRAMF}}
              }
\institute{
    Universit\'e Paris-Saclay, Universit\'e Paris Cit\'e, CEA, CNRS, AIM, 91191, Gif-sur-Yvette, France
    \label{CEA}
    \and
    Institut d'Astrophysique Spatiale (IAS), CNRS, Universit\'e Paris Sud, Orsay, France
    \label{IAS}
    \and
    Institute for Research in Fundamental Sciences (IPM), School of Astronomy, Tehran, Iran
    \label{IPM}
    \and
    Universit\'e C\^ote d'Azur, Observatoire de la C\^ote d'Azur, CNRS, Laboratoire Lagrange, France 
    \label{OCA}
    \and
    School of Physics and Astronomy, Cardiff University, CF24 3AA, UK
    \label{Cardiff}
    \and
    Universit\'e Grenoble Alpes, CNRS, Grenoble INP, LPSC-IN2P3, 38000 Grenoble, France
    \label{LPSC}
    \and	
    Max Planck Institute for Extraterrestrial Physics, 85748 Garching, Germany
    \label{Garching}
    \and
    Aix Marseille Univ, CNRS, CNES, LAM, Marseille, France
    \label{LAM}
    \and
    Universit\'e Grenoble Alpes, CNRS, Institut N\'eel, France
    \label{Neel}
    \and
    Institut de RadioAstronomie Millim\'etrique (IRAM), Grenoble, France
    \label{IRAMF}
    \and 
    Dipartimento di Fisica, Sapienza Universit\`a di Roma, I-00185 Roma, Italy
    \label{Roma}
    \and
    Univ. Grenoble Alpes, CNRS, IPAG, 38000 Grenoble, France
    \label{IPAG}
    \and
    Centro de Astrobiolog\'ia (CSIC-INTA), Torrej\'on de Ardoz, 28850 Madrid, Spain
    \label{CAB}
    \and
    Institute for Research in Fundamental Sciences (IPM), School of Astronomy, Tehran, Iran
    \label{Tehran}
    \and
    National Observatory of Athens, IAASARS, GR-15236, Athens, Greece
    \label{Athens_obs}
    \and
    Faculty of Physics, University of Athens, GR-15784 Zografos, Athens, Greece
    \label{Athens_univ}
    \and
    High Energy Physics Division, Argonne National Laboratory, Lemont, IL 60439, USA
    \label{Argonne}
    \and  
    Instituto de Radioastronom\'ia Milim\'etrica (IRAM), Granada, Spain
    \label{IRAME}
    \and
    LERMA, Observatoire de Paris, PSL Research Univ., CNRS, Sorbonne Univ., UPMC, 75014 Paris, France  
    \label{LERMA}
    \and
    School of Earth \& Space and Department of Physics, Arizona State University, AZ 85287, USA
    \label{Arizona}
    \and
    School of Physics and Astronomy, University of Leeds, Leeds LS2 9JT, UK
    \label{Leeds}
    \and
    INAF-Osservatorio Astronomico di Cagliari, 09047 Selargius, Italy
    \label{INAF}
    \and 
    LPENS, ENS, PSL Research Univ., CNRS, Sorbonne Univ., Universit\'e de Paris, 75005 Paris, France 
    \label{ENS}
    \and  
    Department of Physics and Astronomy, University of Pennsylvania, PA 19104, USA
    \label{Pennsylvanie}
    \and
    Institut d'Astrophysique de Paris, CNRS (UMR7095), 75014 Paris, France
    \label{IAP}
    \and
    University of Lyon, UCB Lyon 1, CNRS/IN2P3, IP2I, 69622 Villeurbanne, France
    \label{IP2I}
    \and
    Sterrenkundig Observatorium Universiteit Gent, Krijgslaan 281 S9, B-9000 Gent, Belgium
    \label{Belguim}
    \and
    IRAP, Universit\'e de Toulouse, CNRS, UPS, IRAP, Toulouse Cedex 4, France
    \label{IRAP}
    }
\abstract{
   We present the first preliminary results of the project \textit{ICED}, focusing on the face-on galaxy NGC4254. 
   We use the millimetre maps observed with NIKA2 at IRAM-30m, as part of the IMEGIN Guaranteed Time Large Program, and of a wide collection of ancillary data (multi-wavelength photometry and gas phase spectral lines) that are publicly available. 
   We derive the global and local properties of interstellar dust grains through infrared-to-radio spectral energy distribution fitting, using the hierarchical Bayesian code HerBIE, which includes the grain properties of the state-of-the-art dust model, THEMIS. 
   Our method allows us to get the following dust parameters: dust mass, average interstellar radiation field, and fraction of small grains. Also, it is effective in retrieving the intrinsic correlations between dust parameters and interstellar medium properties. 
   We find an evident anti-correlation between the interstellar radiation field and the fraction of small grains in the centre of NGC4254, meaning that, at strong radiation field intensities, very small amorphous carbon grains are efficiently destroyed by the ultra-violet photons coming from newly formed stars, through photo-desorption and sublimation. 
   We observe a flattening of the anti-correlation at larger radial distances, which may be driven by the steep metallicity gradient measured in NGC4254.
   }
\maketitle
\section{Introduction}\label{intro}
Interstellar dust (ISD) grains are ubiquitous in galaxies and rather uniformly mixed with the gas in the Interstellar Medium (ISM) \cite{Galliano:2018ARAA}. Although accounting just for $\sim1$\% of the ISM mass \cite{Whittet:book2022}, they have a strong impact on the shape of the Spectral Energy Distribution (SED) of galaxies through the \textit{extinction} of stellar light (i.e. a combination of absorption and scattering), that they re-radiate thermally in the far-infrared (FIR) regime \cite{Whittet:book2022}. 
Most of our knowledge of ISD grain properties comes from studies of the Milky Way (MW) \cite{Draine:2003ARAA}. However, the MW is limited by a narrow range of environmental conditions (e.g. absence of extremely luminous star-forming regions; narrow radial metallicity gradient; passive central black hole) and by confusion along the sightline.
As a consequence, nearby galaxies (i.e., $d <100$ Mpc) have become essential to constrain dust grain properties in extreme conditions and constitute a necessary intermediate step towards understanding distant galaxies \cite{Galliano:2018ARAA}. 
This provides the scientific rationale for the project \textit{ICED}. 
The main objective is to investigate the variation of dust grain properties in response to the different local conditions of the ISM that are observed in nearby galaxies. Eventually, this will allow us to put new constraints on dust models and get further insights into the dust grain lifecycle in the ISM.

In the following, we present a pilot study on NGC4254, a face-on normal galaxy at $d = 12.88$ Mpc, with $D_{25}\simeq5^{\prime}$ \cite{Clark:2018AA}. We use the 1.15 mm and 2 mm maps of NGC4254 by the New IRAM Kid Array (NIKA2; IRAM 30-m telescope) at an angular resolution of $12^{\prime\prime}$ ($\sim2$ kpc) and $18^{\prime\prime}$ ($\sim3$ kpc), respectively \cite{NIKA2-general} \cite{NIKA2-instrument}. They were realized as part of the IMEGIN Guaranteed Time Large program (Interpreting the Millimetre Emission of Galaxies with IRAM-NIKA2; PI: S. Madden), targeting 22 nearby galaxies with diverse ranges of masses, morphologies, metallicities and Star Formation Rates (SFR), for a total of 200 hours of observing time. 
NIKA2 maps are crucial for properly sampling dust thermal SED and constraining the properties of cold dust grains (e.g., dust emissivity index, dust mass and temperature; e.g. \cite{Ejlali:inprep2023}). They are essential to distinguish dust emission from free-free and synchrotron radiation and to investigate the origin of the so-called sub-millimetre excess in galaxies \cite{Katsioli:AA2023}.

\section{Data}
\label{sec-data}
\begin{figure}[!h]
\centering
\includegraphics[scale=0.37]{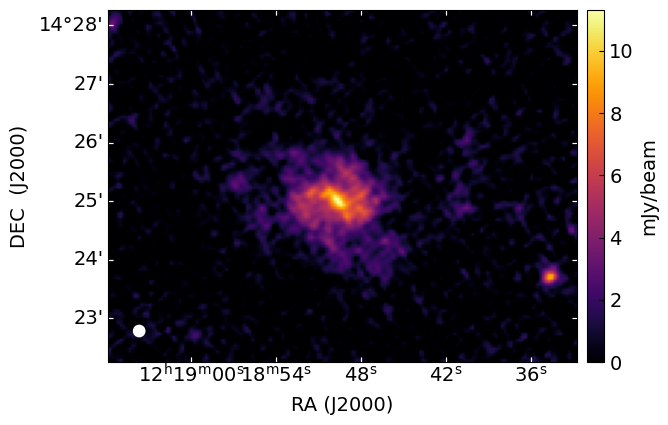}
\includegraphics[scale=0.37]{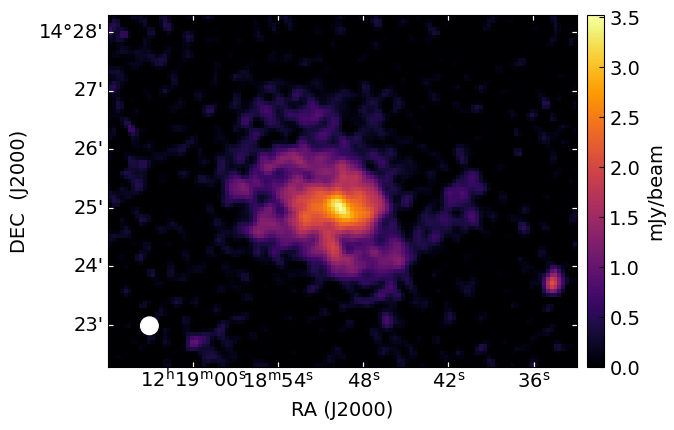}
\caption{NIKA2 maps of NGC4254 at 1.15 mm (left panel; rms $\sim0.8$ mJy/beam) and 2 mm (right panel; rms $\sim0.24$ mJy/beam). Beam size of $12^{\prime\prime}$ ($\sim2$ kpc) and $18^{\prime\prime}$ ($\sim3$ kpc), respectively (cf. white circles in the bottom-left corner).}
\label{fig-NIKA2_maps}       
\end{figure}

We observed NGC4254 with the NIKA2 camera for a total of 24 hours.
The data were reduced using the PIIC pipeline \cite{Zylka:2013} and are shown in Fig.~\ref{fig-NIKA2_maps}.
The ancillary maps used in this work, that we collected from the Dustpedia archive\footnote{http://dustpedia.astro.noa.gr}, include: Spitzer/IRAC at 3.6, 4.5, 5.8, 8.0 $\mu$m (FWHM$\,\sim$ 1.66, 1.72, 1.88, $1.98^{\prime\prime}$); Spitzer/MIPS at 24 $\mu$m (FWHM$\,\sim6^{\prime\prime}$); Hershel/PACS at 70, 100, 160 $\mu$m (FWHM$\,\sim$ 9, 10, $13^{\prime\prime}$); Herschel/SPIRE at 250, 350, 500 $\mu$m (FWHM$\,\sim$ 18, 25, $36^{\prime\prime}$) and Planck at 1380 $\mu$m (FWHM$\,\sim5^{\prime}$). 
We use the combined VLA and Effelsberg radio maps of NGC4254 at 3 cm and 6 cm by \cite{Chyzy:2008} at a final angular resolution of $15^{\prime\prime}$, the CO(1-0) intensity map from the EMPIRE survey (FWHM$\,\sim26^{\prime\prime}$) \cite{EMPIRE:2019}, the CO(2-1) intensity map from the HERACLES survey (FWHM$\,\sim13^{\prime\prime}$) \cite{HERACLES:2009} and the HI intensity map from the VIVA survey (FWHM$\,\sim30^{\prime\prime}$) \cite{VIVA:2009}. 

\section{Method}
We correct the NIKA2 maps for large-scale filtering, i.e. a side consequence of removing the atmospheric noise on large angular scales.
We use the Planck/HFI$_{1380}$ map, which does not suffer from spatial filtering, for measuring the actual global flux of NGC4254. We implement a Monte Carlo algorithm \cite{book:press2007} to extrapolate the integrated flux to 1.15 mm and 2 mm as $\propto \lambda^{2+\beta}$: at each iteration, the dust spectral index is randomly chosen in the range $1<\beta<2$. The amount of flux lost is then redistributed uniformly on each pixel of the galaxy disc. As a result, at 1.15 mm we add back $\sim 3.7\times10^{-5}$ Jy/px, i.e. a total of $\sim 0.45$ Jy ($\sim66$\%); at 2 mm we add back $\sim 5.5\times10^{-6}$ Jy/px, i.e. a total of $\sim 0.04$ Jy ($\sim38$\%). The restored total flux is $0.75\pm0.03$ Jy at 1.15 mm and $0.114\pm0.007$ Jy at 2 mm.
We quantify and subtract the contamination from CO(2-1) line emission from the NIKA2 map at 1.15 mm, both globally and pixel-by-pixel, following \cite{drabek2012}. 
The total contribution of the CO(2-1) line emission to the observed NIKA2 continuum emission at 1.15 mm is $\sim$ 8\%. The CO subtracted integrated flux at 1.15 mm is $0.69\pm0.02$ Jy.

For our purpose, i.e. the spatially-resolved study of dust properties through SED fitting, we need to homogenize the multi-wavelength images of NGC4254, since they originally have different sizes, spatial resolutions, pixel size, orientation, and units. We use the Homogenization of IMEGIN Photometry (HIP) post-processing pipeline \cite{Pantoni:inprep2023} for masking the most luminous foreground stars (mag$_{AB}(J) <12.5$); modelling the foreground/background large-scale emission possibly contaminating the galaxy maps (e.g. Galactic cirrus in the MIR-submm, sky brightness in the NIR, unresolved background sources, any instrumental gradient); convolving to the same angular resolution (SPIRE 500 resolution, i.e. $36^{\prime\prime}$); repixeling and reprojecting to the same pixel size ($12^{\prime\prime}$, i.e. 1/3 of the angular resolution) and orientation. Uncertainties of the original images are propagated through the homogenization procedure, using the bootstrapping Monte Carlo method \cite{book:press2007}.
We fit both the integrated and spatially-resolved (i.e. pixel-by-pixel) dust SED of NGC4254 using the hierarchical Bayesian SED fitting code HerBIE \cite{Galliano:MNRAS2018}, which incorporates the state-of-the-art dust model THEMIS \cite{Jones:AAP2017}. We model the dust emission using the non-uniformly illuminated dust mixture, as it is implemented in HerBIE (free parameters: dust mass, $M_{dust}$; minimum ISRF intensity, $U_m$; fraction of aromatic feature carriers, $q_{AF}$). We also include the NIR emission by stellar populations (modelled as a black body of given temperature) and free–free and synchrotron continua. HerBIE returns the probability density functions and the map of dust parameters along with their uncertainties (cf. Fig.~\ref{fig-dust_par_maps}).

\section{Results and discussion}
\begin{figure}[!t]
\centering
\sidecaption
\includegraphics[width=6cm,clip]{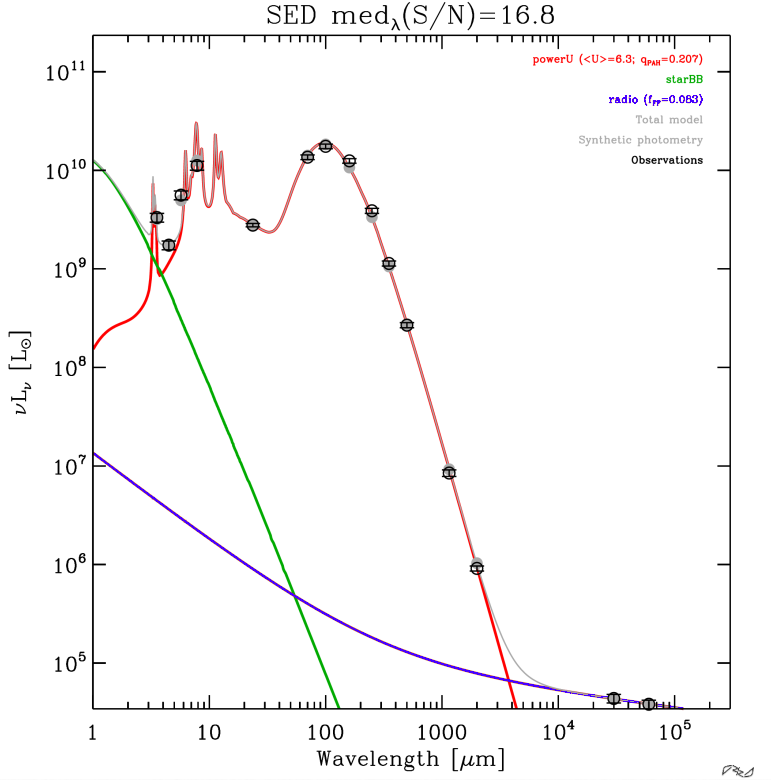}
\caption{Global NIR-to-radio SED fitting of NGC4254 with HerBIE. The red line shows the resulting ISD emission; the green line shows the modelled stellar emission; the blue line shows the modelled radio emission (synchrotron and free-free); the light grey line is the sum of all the previous components (total model). Black empty circles are the data points, while grey-filled circles are the values expected from the modelling. NIKA2 global photometry at 1.15 mm and 2 mm is corrected for large-scale spatial filtering. CO(2-1) line contribution is subtracted from the 1.15 mm map. }
\label{fig-global_SED}
\end{figure}

\begin{figure}[!h]
\centering
\includegraphics[scale=0.25]{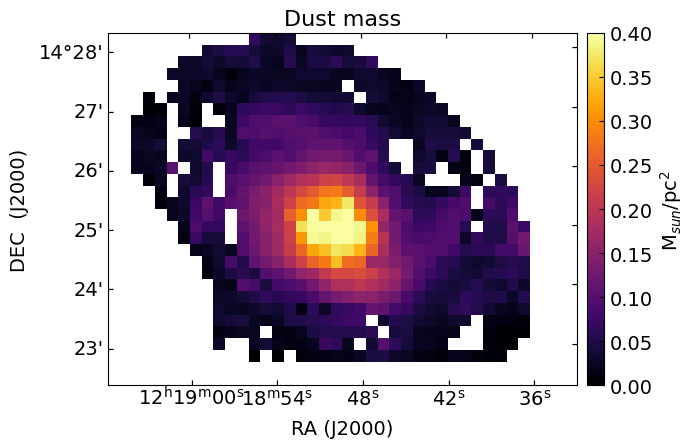}
\includegraphics[scale=0.25]{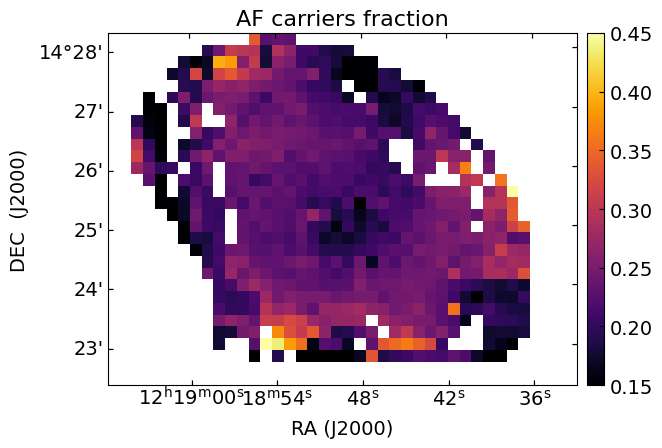}
\includegraphics[scale=0.25]{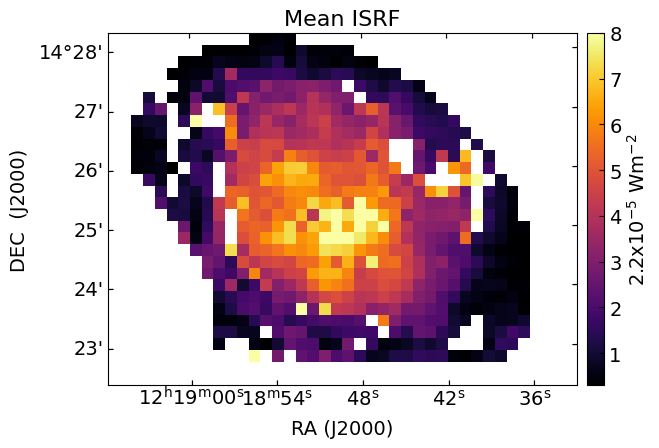}
\caption{HerBIE dust parameter maps of NGC4254: dust mass map (left);
fraction of aromatic feature carries, $q_{AF}$ (centre); average ISRF intensity, $\langle U \rangle$
(right). }
\label{fig-dust_par_maps}
\end{figure}

\begin{figure}[!h]
\centering
\sidecaption
\includegraphics[width=7cm,clip]{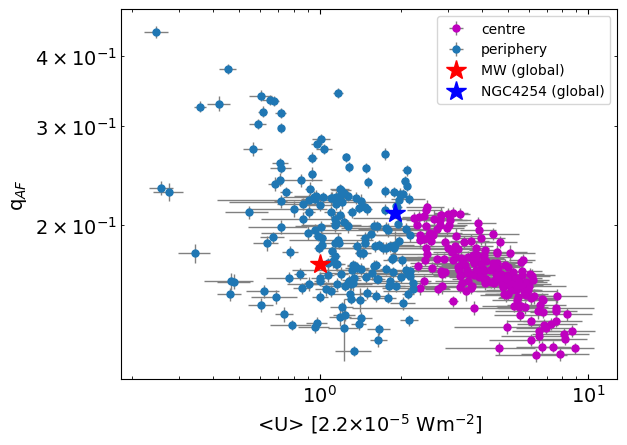}
\caption{Spatially resolved anti-correlation between the fraction of aromatic feature carries, $q_{AF}$, and the average ISRF intensity, $\langle U \rangle$. Each filled circle stands for one pixel ($12^{\prime\prime}$) in the galaxy maps. Magenta circles represent the pixels in the galaxy centre, while the light blue circles identify the galaxy outskirts. Red and blue stars show the global values for MW and NGC4254, respectively.}
\label{fig-anti-corr}
\end{figure}

 We measured the integrated photometry of NGC4254 within an ellipse centred at [RA 184.70655 deg; DEC 14.41641 deg], with a semi-major axis of $220^{\prime\prime}$ and an axial ratio of 1.3471 \cite{Clark:2018AA}, and we checked it against literature. The global SED fit is displayed in Fig.~\ref{fig-global_SED}. We include all the bands listed in Sect.~\ref{sec-data}, except for Planck/HFI$_{1380}$.
 The SED fitting gives us a total dust mass $M_{dust}=(2.5\pm0.1)\times10^7$ M$_\odot$, where the small grains fraction is $q_{AF}=0.21\pm0.01$ (cf. Fig.~\ref{fig-anti-corr}). We use the CO(1-0) and HI intensity maps for measuring the total molecular and atomic gas content: $M_{mol}= (5\pm1)\times10^9$ M$_\odot$ 
 and $M_{HI}= (6.5\pm0.7)\times10^9$ M$_\odot$
 \cite{Casasola:2017}. Using the calibrations given in \cite{Leroy:2008}, we derive the total SFR 
$=2.4\pm0.5$ M$_\odot$ yr$^{-1}$ and stellar mass $M_{star}=
(3.4\pm0.7)\times10^{10}$ M$_\odot$. These values are consistent with previous studies on NGC4254, e.g. \cite{Chemin:2016}. 
 The pixel-by-pixel SED fitting provides us with the maps of three ISD parameters (i.e. $M_{dust}$, $q_{AF}$ and $\langle U \rangle$), that are displayed in Fig.~\ref{fig-dust_par_maps}. The dust mass, tracing the distribution of large grains in the galaxy, is mostly located in the central part of the NGC4254 and along the two most prominent spiral arms of the galaxy. Small grains carrying aromatic features are located preferentially in the periphery of NGC4254, leaving a clear hole in the galaxy centre.
 The averaged ISRF intensity peaks in the centre of NGC4254 and progressively decreases towards the periphery, showing a morphology almost complementary to the small dust grain fraction.
The scatter plot of the $q_{AF}$ and $\langle U \rangle$ maps (Fig.~\ref{fig-anti-corr}) shows the existence of an anti-correlation between these two quantities, that is evident in the central pixels of the galaxy (i.e., within $\sim3.5^{\prime}$; cf. magenta circles). This indicates that, in strong radiation field conditions, small dust grains are very efficiently depleted: the strong UV radiation from newly formed stars destroys the smallest dust grains through photo-desorption and sublimation.
Even if the scatter is quite large, there is evidence of a flattening towards the galaxy periphery (cf. light blue circles in Fig.~\ref{fig-anti-corr}). We believe this trend to be driven by the steep metallicity gradient measured in NGC4254 \cite{Boselli:2018}. However, it could also arise from a combination of other processes, such as coagulation of small grains onto larger grains; lack of UV photons exciting the grains in passive molecular clouds; or thermal sputtering destroying the small grains in the diffuse ionised gas. Further analysis is needed to confirm the scenario, such as a proper comparison with the local physical conditions of the gas phase.
 
\section{Summary}
We presented and discussed the first preliminary results of the project \textit{ICED}, focusing on the face-on galaxy NGC4254. 
For constraining ISD grains and ISM properties in NGC4254, we took advantage of the NIKA2 (IRAM-30m) maps at 1.15 mm and 2 mm (IMEGIN Guaranteed Time Large Program) and of a wide collection of ancillary data from public catalogues. We corrected for the large-scale emission filtering affecting the NIKA2 maps and we subtracted the contribution from CO(2-1) to the millimetre continuum. We homogenized the data maps and propagated uncertainties via Monte Carlo method. We derived dust grain global and local properties by fitting the NIR-to-radio SED of NGC4254 using the hierarchical Bayesian fitting code HerBIE, which includes the state-of-the-art dust model THEMIS. 
Our method allowed us to get the most important dust parameter maps (e.g., dust mass, ISRF, fraction of small grains) and it was effective in retrieving the intrinsic correlations between dust parameters and ISM properties. 
We found an evident anti-correlation between the average ISRF intensity and the fraction of small grains carrying the aromatic features in the galaxy centre. We concluded that, in very strong radiation field conditions (higher than twice the ISRF intensity in the solar neighbourhood), small grains are very efficiently destroyed by UV photons from newly formed stars (through photo-desorption and sublimation). 
The flattening of the correlation at larger radial distances may be traced back to the steep metallicity gradient measured in NGC4254. Further analysis is needed to confirm this scenario.
%
\section*{Acknowledgements}
\footnotesize{The NIKA2 data were processed using the Pointing and Imaging In Continuum (PIIC) software, developed by Robert Zylka at IRAM and distributed by IRAM via the GILDAS pages. PIIC is the extension of the MOPSIC data reduction software to the case of NIKA2 data. We would like to thank the IRAM staff for their support during the observation campaigns. The NIKA2 dilution cryostat has been designed and built at the Institut N\'eel. In particular, we acknowledge the crucial contribution of the Cryogenics Group, and in particular Gregory Garde, Henri Rodenas, Jean-Paul Leggeri, Philippe Camus. This work has been partially funded by the Foundation Nanoscience Grenoble and the LabEx FOCUS ANR-11-LABX-0013. This work is supported by the French National Research Agency under the contracts "MKIDS", "NIKA" and ANR-15-CE31-0017 and in the framework of the "Investissements d’avenir” program (ANR-15-IDEX-02). This work has benefited from the support of the European Research Council Advanced Grant ORISTARS under the European Union's Seventh Framework Programme (Grant Agreement no. 291294). E. A. acknowledges funding from the French Programme d’investissements d’avenir through the Enigmass Labex. A. R. acknowledges financial support from the Italian Ministry of University and Research - Project Proposal CIR01$\_00010$. M.B., A.N., and S.C.M. acknowledge support from the Flemish Fund for Scientific Research (FWO-Vlaanderen, research project G0C4723N). S. K. acknowledges support provided by the Hellenic Foundation for Research and Innovation (HFRI) under the 3rd Call for HFRI PhD Fellowships (Fellowship Number: 5357). This work was funded by the P2IO LabEx (ANR-10-LABX-0038) in the framework "Investissements d'Avenir" (ANR-11-IDEX-0003-01) managed by the Agence Nationale de la Recherche (ANR, France). This work was supported by the Programme National Physique et Chimie du Milieu Interstellaire (PCMI) and the Programme National Cosmology et Galaxies (PNCG) of the CNRS/INSU with INC/INP co-funded by CEA and CNES.} 

\end{document}